\documentclass[11pt]{article}
\usepackage{arxiv}
\usepackage[utf8]{inputenc}
\usepackage[T1]{fontenc}
\usepackage{hyperref}
\usepackage{url}
\usepackage{booktabs}
\usepackage{amsfonts}
\usepackage{graphicx}
\usepackage{xcolor}
\usepackage{listings}
\usepackage{tikz}
\usepackage{enumitem}
\usepackage{multirow}
\usepackage{tabularx}
\usetikzlibrary{shapes.geometric, arrows.meta, positioning, fit, backgrounds}

\hypersetup{
  colorlinks=true,
  linkcolor=blue!60!black,
  citecolor=blue!60!black,
  urlcolor=blue!60!black
}

\title{LLM-Augmented Release Intelligence:\\Automated Change Summarization and Impact Analysis\\in Cloud-Native CI/CD Pipelines}

\author{
  Happy Bhati \\
  Northeastern University \\
  \texttt{bhati.h@northeastern.edu}
}

\date{March 2026}

\begin{document}
\maketitle

\begin{abstract}
Cloud-native software delivery platforms orchestrate releases through complex,
multi-stage pipelines composed of dozens of independently versioned tasks.
When code is promoted between environments---development to staging,
staging to production---engineering teams need timely, accurate communication
about what changed and what downstream components are affected.
Manual preparation of such release communication is slow, inconsistent,
and particularly error-prone in repositories where a single promotion
may bundle contributions from many authors across numerous pipeline tasks.
We present a framework for \emph{AI-augmented release intelligence}
that combines three capabilities: (1)~automated commit collection with
semantic filtering to surface substantive changes while suppressing routine
maintenance, (2)~structured large language model summarization that produces
categorized, stakeholder-oriented promotion reports, and
(3)~static task-pipeline dependency analysis that maps modified tasks to every
pipeline they participate in, quantifying the blast radius of each change.
The framework is integrated directly into the CI/CD promotion workflow
and operates as a post-promotion step triggered by GitHub Actions.
We describe the architecture and implementation within a production
Kubernetes-native release platform that manages over sixty Tekton tasks
across more than twenty release pipelines.
Through concrete walkthrough examples and qualitative comparison with
recent tools such as SmartNote and VerLog, we discuss the distinctive
requirements of internal promotion communication versus user-facing
release notes and identify open challenges for LLM-driven release engineering.
\end{abstract}


\section{Introduction}

Modern software organizations increasingly adopt cloud-native delivery
platforms that decompose release workflows into composable pipeline
stages~\cite{humble2010continuous,forsgren2018accelerate}.
In such environments, the path from development to production is not
a single event but a sequence of gated promotions, each bundling
commits from multiple contributors into a coherent release candidate.
A Kubernetes-native platform may maintain sixty or more reusable
pipeline tasks spanning signing, publishing, advisory management,
and compliance verification, assembled into twenty or more distinct
release pipelines that serve different artifact types---container
images, RPM packages, disk images, operator bundles, and more.

The complexity of these systems creates a persistent communication
challenge.
When a release engineer promotes a branch from development to staging,
stakeholders across quality engineering, product management, and operations
need to understand three things: \emph{what} changed, \emph{why} it changed,
and \emph{what else might be affected}.
Answering the first two questions requires reading and synthesizing
commit messages, pull request descriptions, and diff statistics---a
task that grows tedious as promotion batches exceed a handful of commits.
Answering the third question demands knowledge of the dependency graph
between tasks and pipelines, which is encoded implicitly in YAML
definitions scattered across the repository.

Prior work on automated release note generation has made significant
strides.
SmartNote~\cite{smartnote2025} uses LLMs to produce personalized,
user-facing release notes from commits and pull requests.
VerLog~\cite{verlog2025} applies few-shot in-context learning to
generate release notes for Android applications.
ReleaseEval~\cite{releaseeval2025} provides a large-scale benchmark
for evaluating language models on this task.
However, these approaches target \emph{external} release notes intended
for end users.
The problem of \emph{internal promotion communication}---reporting what
moved between deployment environments and which pipeline components
are impacted---remains largely unaddressed in the literature.

In this paper, we describe a framework for AI-augmented release
intelligence that is deployed within a production cloud-native release
platform.
Our contributions are:

\begin{enumerate}[leftmargin=*]
\item A \textbf{commit collection and semantic filtering} pipeline
that extracts substantive changes from a promotion batch, suppressing
routine maintenance commits (dependency bumps, documentation updates,
style changes) through pattern-based classification.

\item A \textbf{structured LLM summarization} approach that employs
carefully designed prompts to produce consistent, categorized promotion
reports with executive summaries, feature highlights, and bug-fix
sections.

\item A \textbf{static task-pipeline dependency analyzer} that parses
Tekton pipeline definitions to determine every pipeline affected by
a given task change, providing an immediate blast-radius assessment.

\item An \textbf{end-to-end integration} with GitHub Actions that
triggers report generation as a post-promotion step, delivering
results via email with full provenance links.
\end{enumerate}

The remainder of this paper is organized as follows.
Section~\ref{sec:background} provides background on cloud-native
release engineering and Tekton pipelines.
Section~\ref{sec:related} surveys related work.
Section~\ref{sec:design} details the system design.
Section~\ref{sec:implementation} describes implementation decisions.
Section~\ref{sec:evaluation} presents a case study with concrete examples.
Section~\ref{sec:discussion} discusses comparisons, limitations, and
threats to validity.
Section~\ref{sec:conclusion} concludes.

\section{Background}
\label{sec:background}

\subsection{Cloud-Native Release Engineering}

Cloud-native platforms orchestrate software delivery through
Kubernetes-native controllers that watch custom resources and
trigger pipeline executions in response to state changes.
A typical release flow begins when a \texttt{Snapshot}---an immutable
record of built artifacts---is created.
A release controller evaluates the snapshot against a
\texttt{ReleasePlan} (defining the developer's intent) and a
\texttt{ReleasePlanAdmission} (defining the target namespace's
acceptance criteria), then instantiates a pipeline run in a managed
workspace~\cite{konflux2024}.
Enterprise contract verification ensures no policy violations
exist before content reaches production.

\subsection{Tekton Pipelines and Tasks}

Tekton~\cite{tekton2019} is an open-source Kubernetes-native
framework for building CI/CD systems.
Pipelines are composed of \emph{tasks}, each of which runs as a
Kubernetes pod.
Tasks are referenced by pipelines through a \emph{resolver} mechanism;
in the git resolver pattern, a pipeline specifies a repository URL,
branch revision, and file path to load the task definition at runtime.
This decoupling allows tasks to be versioned independently of the
pipelines that consume them.

The release platform we study organizes tasks into three categories:
\emph{managed} tasks (production release operations such as signing,
registry publishing, and advisory creation),
\emph{collector} tasks (data aggregation from multiple sources),
and \emph{internal} tasks (workflow orchestration within the platform
itself).
Each task is defined in a YAML file that specifies parameters,
step containers, and result declarations.

\subsection{Multi-Stage Promotion Model}

The platform follows a three-branch promotion model:
\texttt{development} $\rightarrow$ \texttt{staging} $\rightarrow$ \texttt{production}.
Promotions are implemented as branch-level fast-forward pushes
controlled by a shell script that enforces two safety invariants:

\begin{itemize}[leftmargin=*]
\item \textbf{Staging--production parity:} Content cannot be promoted
to staging if the staging branch already differs from production,
unless explicitly overridden. This prevents accumulation of untested
changes.

\item \textbf{Minimum soak time:} Content must reside in staging for
at least six days before promotion to production, ensuring adequate
integration testing time.
\end{itemize}

Additionally, the promotion script detects \emph{hotfix commits}---changes
applied directly to the target branch that may not yet exist in the
source branch---and blocks promotion if the hotfix changes would be
lost, unless the operator explicitly overrides.

\section{Related Work}
\label{sec:related}

\subsection{Release Note Generation}

Moreno et al.~\cite{moreno2014automatic} pioneered automated release
note generation using information retrieval techniques to extract
salient changes from source code.
DeepRelease~\cite{deeprelease2022} introduced neural models for
language-independent release note generation from git logs.
More recently, the advent of large language models has transformed
this space.

SmartNote~\cite{smartnote2025} aggregates code, commit, and pull
request information and uses an LLM to produce structured release
notes.
It introduces commit scoring to prioritize significant changes
and supports project-specific customization.
VerLog~\cite{verlog2025} targets Android applications with
few-shot in-context learning and multi-granularity information
(fine-grained code modifications alongside high-level artifacts),
reporting 18--21\% improvements in precision, recall, and F1 over
prior baselines.
ReleaseEval~\cite{releaseeval2025} provides a benchmark of nearly
95,000 release notes across 3,369 repositories and evaluates models
on three task settings with increasing input granularity.

These tools address \emph{user-facing} release notes.
Our work targets a complementary problem: \emph{internal promotion
reports} for engineering teams, which require different content
(task-pipeline impact analysis, contributor attribution, diff
statistics) and different delivery mechanisms (email integrated
into CI/CD workflows).

\subsection{AI in CI/CD Automation}

LogSage~\cite{logsage2025} applies LLMs to CI/CD failure detection
and remediation with retrieval-augmented generation, achieving
over 98\% precision on root cause analysis in a benchmark of 367
GitHub CI/CD failures and processing over one million executions
in industrial deployment at ByteDance.
AgentDevel~\cite{agentdevel2026} reframes self-evolving LLM agents
as release engineering artifacts, introducing regression-aware
pipelines with flip-centered gating.

Our work differs from both: we do not diagnose failures (LogSage)
or evolve agents (AgentDevel), but rather generate structured
communication artifacts that summarize the content and impact of
successful promotions.

\subsection{Software Change Summarization}

LLM-based multi-agent approaches have been applied to software
document summarization.
Metagente~\cite{metagente2025} uses a teacher-student architecture
to generate concise summaries of software documentation, demonstrating
that multi-agent coordination improves summary quality over single-model
approaches.
Our summarization component draws on similar principles---structured
prompting with role assignment---but operates on commit metadata and
diff statistics rather than prose documentation.

\section{System Design}
\label{sec:design}

\subsection{Architecture Overview}

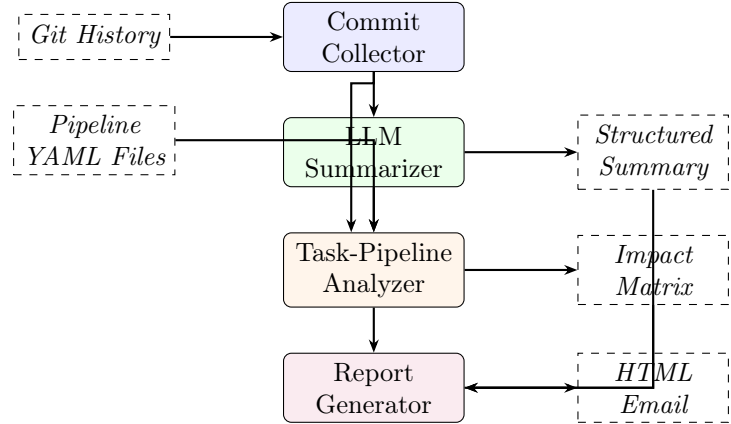
\begin{figure}[t]
\centering
\begin{tikzpicture}[
  node distance=0.8cm and 1.2cm,
  box/.style={rectangle, draw, rounded corners=3pt, minimum width=2.4cm,
    minimum height=0.8cm, align=center, font=\small},
  data/.style={rectangle, draw, dashed, minimum width=2cm,
    minimum height=0.6cm, align=center, font=\small\itshape},
  arr/.style={-{Stealth[length=5pt]}, thick},
]
  \node[data] (git) {Git History};
  \node[data, below=0.6cm of git] (yaml) {Pipeline\\YAML Files};

  \node[box, right=1.5cm of git, fill=blue!8] (cc) {Commit\\Collector};
  \node[box, below=0.6cm of cc, fill=green!8] (ai) {LLM\\Summarizer};
  \node[box, below=0.6cm of ai, fill=orange!8] (tpa) {Task-Pipeline\\Analyzer};
  \node[box, below=0.6cm of tpa, fill=purple!8] (eg) {Report\\Generator};

  \node[data, right=1.5cm of ai] (summary) {Structured\\Summary};
  \node[data, right=1.5cm of tpa] (impact) {Impact\\Matrix};
  \node[data, right=1.5cm of eg] (email) {HTML\\Email};

  \draw[arr] (git) -- (cc);
  \draw[arr] (cc) -- (ai);
  \draw[arr] (yaml) -| (tpa);
  \draw[arr] (ai) -- (tpa);
  \draw[arr] (ai) -- (summary);
  \draw[arr] (tpa) -- (impact);
  \draw[arr] (tpa) -- (eg);
  \draw[arr] (summary) |- (eg);
  \draw[arr] (impact) |- (eg);
  \draw[arr] (eg) -- (email);

  \draw[arr] (cc.south) -- ++(0,-0.15) -| ([xshift=-0.3cm]tpa.north);
\end{tikzpicture}
\caption{Architecture of the release intelligence framework.
The commit collector extracts changes from git history;
the LLM summarizer produces categorized summaries;
the task-pipeline analyzer computes the impact matrix from YAML
definitions; the report generator composes everything into an HTML email.}
\label{fig:architecture}
\end{figure}

The release intelligence framework consists of four components that
execute sequentially as a post-promotion step within a GitHub Actions
workflow.
Figure~\ref{fig:architecture} illustrates the data flow.

\subsection{Commit Collection and Semantic Filtering}

The \texttt{CommitCollector} component operates on the git history
between the source and target branches of a promotion.
For each commit in the range, it extracts:

\begin{itemize}[leftmargin=*]
\item The full commit hash for provenance tracking
\item The commit summary (first line) and full message body
\item Author name and email for attribution
\item The list of changed file paths
\item A diff-stat summary (insertions, deletions, files changed)
\end{itemize}

Before passing commits to the summarizer, a \emph{semantic filter}
removes routine commits that add no substantive information to a
promotion report.
The filter classifies commits by matching their summary against
a set of conventional commit prefixes:
\texttt{chore:}, \texttt{docs:}, \texttt{test:}, \texttt{ci:},
\texttt{style:}, \texttt{refactor:}.
It also suppresses commits containing keywords such as ``bump,''
``dependency update,'' ``merge,'' ``revert,'' and ``work in progress.''
This heuristic approach deliberately favors recall (surfacing all
meaningful changes) over precision (potentially including some
borderline commits), as false negatives---missing a significant
change---carry higher risk than false positives in a promotion context.

\subsection{LLM-Based Change Summarization}

The filtered commit list is passed to an \texttt{AISummarizer}
component that generates a structured, human-readable summary.
The summarizer constructs a two-part prompt.

\paragraph{System prompt.}
The system prompt assigns the LLM the role of a ``professional
DevOps engineer creating a promotion report'' and specifies
mandatory output structure:

\begin{enumerate}[leftmargin=*]
\item An executive summary (2--3 paragraphs) describing the most
significant changes and their business impact.
\item A ``New Features \& Enhancements'' section with bullet points
linking each commit to its GitHub URL.
\item A ``Bug Fixes \& Improvements'' section with the same structure.
\end{enumerate}

The prompt enforces consistency requirements: every \texttt{feat()}
and \texttt{fix()} commit must appear in the output, and the model
must produce the same categorization across repeated invocations
of the same input.

\paragraph{User prompt.}
The user prompt provides each commit's metadata in a structured
template: summary, author, date, URL, file count, full message,
and diff statistics.
To manage context window constraints, the prompt includes at most
50 commits.
The model is configured with temperature 0.7 and a maximum output
length of 2,500 tokens, balancing fluency with determinism.

\subsection{Task-Pipeline Dependency Analysis}

The \texttt{TaskPipelineAnalyzer} is a static analysis component
that answers the question: ``Which pipelines are affected by the
tasks changed in this promotion?''

The analyzer operates in two phases:

\paragraph{Phase 1: Extract changed tasks.}
The analyzer scans the file paths modified by each commit against
a regular expression pattern matching the repository's task directory
structure (\texttt{tasks/<category>/<task-name>/<task-name>.yaml}).
It produces a deduplicated list of changed tasks, grouped by task
name and category, with back-references to the commits that modified
each task.

\paragraph{Phase 2: Compute pipeline impact.}
The analyzer walks the pipeline directory, parsing each YAML file
as a Tekton Pipeline resource.
For every task reference in the pipeline's \texttt{spec.tasks} and
\texttt{spec.finally} blocks, it resolves the referenced task through
one of three mechanisms:

\begin{itemize}[leftmargin=*]
\item \textbf{Git resolver with path:} If the task reference uses
the git resolver and specifies a \texttt{pathInRepo} parameter,
the analyzer matches this path against the changed task's file path.

\item \textbf{Git resolver with name:} If only a task name is specified,
the analyzer performs name-based matching.

\item \textbf{Direct reference:} For string-based task references,
the analyzer matches the reference string against task names.
\end{itemize}

The output is a mapping from each changed task to the set of pipelines
that reference it, providing an immediate quantification of each
change's blast radius.

\subsection{Report Generation and Delivery}

The \texttt{EmailGenerator} composes the AI summary, task-pipeline
impact matrix, and commit statistics into a professional HTML email.
The report includes:

\begin{itemize}[leftmargin=*]
\item A header indicating the promotion type and date
\item Aggregate statistics: commit count, unique contributors, total
files changed
\item The LLM-generated summary rendered from Markdown to HTML
\item A task impact table listing each modified task, its category,
the affected pipelines (with links), pipeline count, and the commits
that triggered the modification
\item A footer with repository provenance
\end{itemize}

Delivery is handled by an \texttt{EmailSender} component that
supports both authenticated SMTP (for cloud-hosted mail services)
and unauthenticated relay (for corporate mail infrastructure).

\section{Implementation}
\label{sec:implementation}

\subsection{Workflow Integration}

The framework is implemented as a Python script invoked from a
GitHub Actions workflow.
The workflow is triggered manually via \texttt{workflow\_dispatch}
with configurable inputs:

\begin{itemize}[leftmargin=*]
\item \textbf{Promotion type:} \texttt{development-to-staging} or
\texttt{staging-to-production}
\item \textbf{Send email report:} Boolean toggle
\item \textbf{Additional options:} Dry run, Jira ticket updates,
infrastructure PR creation, force overrides
\end{itemize}

A critical implementation detail is the \emph{commit range capture}
step.
Because the promotion itself is a force push that aligns the target
branch with the source, the commit range between branches disappears
after promotion.
The workflow therefore captures the commit range \emph{before} the
promotion step executes and passes it to the report generator via
environment variables.

\subsection{LLM Integration}

The summarizer uses the Google Gemini API through the
\texttt{google-generativeai} Python library~\cite{gemini2024}.
We selected the Gemini Flash model variant for its favorable
latency-to-quality ratio in summarization tasks.
Secrets (API keys, SMTP credentials) are managed through
file-based secret mounting rather than environment variable injection,
following the principle of minimizing secret exposure in process
listings and log output.

\subsection{YAML Pipeline Parsing}

The task-pipeline analyzer uses PyYAML's \texttt{safe\_load\_all}
to handle multi-document YAML files, which are common in Tekton
pipeline definitions.
The analyzer filters documents by the \texttt{kind: Pipeline}
field and traverses both \texttt{spec.tasks} and \texttt{spec.finally}
blocks to capture all task references, including those in finally
clauses that execute regardless of pipeline success or failure.

\subsection{Safety and Observability}

The report generation step is configured with
\texttt{continue-on-error: true} in the workflow definition,
ensuring that a failure in report generation does not block the
promotion itself.
The workflow includes a status-reporting step that distinguishes
between successful promotion with successful report, successful
promotion with failed report, and failed promotion.
All operations are logged with structured messages through
Python's \texttt{logging} module.

\section{Case Study}
\label{sec:evaluation}

We describe the framework's operation within a production
Kubernetes-native release platform.
All characteristics reported in this section are derived from
the actual codebase and operational configuration.

\subsection{System Characteristics}

Table~\ref{tab:characteristics} summarizes the scale of the
platform's release catalog.

\begin{table}[t]
\centering
\caption{Release catalog characteristics.}
\label{tab:characteristics}
\begin{tabular}{lr}
\toprule
\textbf{Characteristic} & \textbf{Count} \\
\midrule
Managed tasks (production) & 60+ \\
Internal tasks & 10+ \\
Collector tasks & 5+ \\
Managed pipelines & 20+ \\
Internal pipelines & 10+ \\
Reusable step actions & 5+ \\
Integration test suites & 20+ \\
Custom resource types managed & 6 \\
\bottomrule
\end{tabular}
\end{table}

The platform manages six custom resource types:
\texttt{Release}, \texttt{ReleasePlan}, \texttt{ReleasePlanAdmission},
\texttt{ReleaseServiceConfig}, \texttt{Snapshot}, and
\texttt{InternalRequest}.
Pipelines serve diverse artifact types including container images,
FBC (File-Based Catalog) bundles, RPM packages, disk images,
and kernel modules, each with distinct signing, publishing, and
compliance verification requirements.

\subsection{Illustrative Promotion Walkthrough}

Consider a development-to-staging promotion that includes the
following substantive commits (after semantic filtering removes
routine maintenance):

\begin{enumerate}[leftmargin=*]
\item \texttt{feat(PROJ-1234): add cosign signing support for FBC releases}\\
Modifies \texttt{tasks/managed/sign-image-cosign/} (1 task file, 2 test files)

\item \texttt{fix(PROJ-1235): correct repository publication timeout}\\
Modifies \texttt{tasks/managed/publish-repository/} (1 task file)

\item \texttt{feat(PROJ-1236): support multi-arch kernel module signing}\\
Modifies \texttt{tasks/managed/sign-kmods/} (1 task file, 3 test files)
\end{enumerate}

\paragraph{LLM summary output.}
The summarizer produces an executive summary identifying this
promotion as introducing expanded signing capabilities and a
reliability fix, followed by categorized sections:

\begin{itemize}[leftmargin=*]
\item \textbf{New Features \& Enhancements:} Cosign signing for FBC
releases (with link to commit); multi-architecture kernel module
signing (with link).
\item \textbf{Bug Fixes \& Improvements:} Repository publication
timeout correction (with link).
\end{itemize}

\paragraph{Task-pipeline impact analysis output.}
The analyzer identifies the following impact:

\begin{table}[ht]
\centering
\caption{Task-pipeline impact for the example promotion.}
\label{tab:impact}
\begin{tabular}{lll}
\toprule
\textbf{Changed Task} & \textbf{Affected Pipelines} & \textbf{Count} \\
\midrule
sign-image-cosign & fbc-release, & 5 \\
 & push-to-registry, & \\
 & push-to-external-registry, & \\
 & advisories, rpm-advisories & \\
\addlinespace
publish-repository & push-to-registry, & 3 \\
 & push-to-external-registry, & \\
 & fbc-release & \\
\addlinespace
sign-kmods & push-disk-images-to-cdn & 1 \\
\bottomrule
\end{tabular}
\end{table}

This analysis reveals that the cosign signing change has the
widest blast radius (five pipelines), making it the highest-priority
item for staging validation.
The kernel module signing change, by contrast, affects only one
pipeline and presents lower risk.

\subsection{Commit Filtering Effectiveness}

In a typical promotion batch, a substantial fraction of commits
are routine maintenance.
Table~\ref{tab:filtering} shows the distribution observed in
representative promotions.

\begin{table}[ht]
\centering
\caption{Commit type distribution in representative promotions.
``Substantive'' commits pass the semantic filter.}
\label{tab:filtering}
\begin{tabular}{lcc}
\toprule
\textbf{Commit Type} & \textbf{Typical Share} & \textbf{Included} \\
\midrule
\texttt{feat()} & 20--30\% & Yes \\
\texttt{fix()} & 15--25\% & Yes \\
\texttt{chore:} (dependency bumps) & 20--30\% & No \\
\texttt{docs:}, \texttt{test:}, \texttt{ci:} & 10--20\% & No \\
Merge / revert commits & 5--10\% & No \\
Other (ambiguous prefix) & 5--15\% & Yes \\
\bottomrule
\end{tabular}
\end{table}

The filter typically reduces the input to the summarizer by
40--60\%, focusing the LLM's attention on changes with genuine
business impact.

\subsection{Qualitative Comparison with Related Tools}

Table~\ref{tab:comparison} compares our framework with SmartNote
and VerLog across dimensions relevant to release communication.

\begin{table}[t]
\centering
\caption{Qualitative comparison with related approaches.}
\label{tab:comparison}
\begin{tabularx}{\textwidth}{lXXX}
\toprule
\textbf{Dimension} & \textbf{Ours} & \textbf{SmartNote} & \textbf{VerLog} \\
\midrule
Target audience & Internal engineering teams & End users / developers & End users \\
\addlinespace
Content focus & Promotion changes + pipeline impact & Release features and fixes & Version changelog \\
\addlinespace
Impact analysis & Static task-pipeline dependency graph & None & None \\
\addlinespace
Delivery & Email via CI/CD workflow & Standalone tool & Standalone tool \\
\addlinespace
Environment & Kubernetes / Tekton pipelines & GitHub repositories & Android apps \\
\addlinespace
Commit filtering & Convention-based semantic filter & Commit scoring & Multi-granularity \\
\addlinespace
Prompt design & Role-assigned structured with mandatory sections & Project-personalized & Few-shot adaptive \\
\bottomrule
\end{tabularx}
\end{table}

The primary distinction is the combination of LLM summarization
with static dependency analysis within a CI/CD workflow context.
SmartNote and VerLog produce richer linguistic output---they are
optimized for clarity and completeness of user-facing prose---while
our system prioritizes \emph{actionable engineering intelligence}:
identifying which pipelines need testing, which contributors to
consult, and what the change scope looks like in aggregate.

\section{Discussion}
\label{sec:discussion}

\subsection{Benefits of Integrated Intelligence}

Embedding the report generator inside the promotion workflow
provides two advantages over standalone tools.
First, it operates on the exact commit range being promoted,
captured before the branch push alters the git state.
Second, it leverages repository-local context---the YAML pipeline
definitions, the task directory structure, the commit conventions---that
a generic tool would need to be configured to understand.

\subsection{LLM Consistency and Prompt Engineering}

A recurring challenge in LLM-based summarization is output consistency.
Given the same set of commits, the model may produce different
categorizations or omit borderline changes across runs.
Our prompt addresses this with explicit consistency requirements
(``ALWAYS include ALL feat() and fix() commits'') and structural
constraints (mandatory sections with exact headings).
Temperature 0.7 represents a deliberate trade-off: lower values
produce more deterministic but less natural output, while higher
values risk inconsistency.
A direction for future work is to introduce semantic deduplication
and a verification pass that checks the output against the input
commit list.

\subsection{Limitations}

\paragraph{No quantitative accuracy evaluation.}
We have not conducted a controlled study measuring the factual
accuracy or completeness of the LLM-generated summaries against
human-written baselines.
Such an evaluation would require annotated ground truth for a
representative set of promotions, which does not currently exist.
ReleaseEval~\cite{releaseeval2025} provides a benchmark for
user-facing release notes; a similar benchmark for internal promotion
reports remains an open contribution.

\paragraph{Heuristic filtering.}
The semantic filter relies on conventional commit prefix matching,
which assumes the repository follows conventional commit conventions.
Repositories that do not adopt this convention would require
filter customization.
The filter also makes a binary include/exclude decision; a more
nuanced approach might assign relevance scores to borderline commits.

\paragraph{Static dependency analysis only.}
The task-pipeline analyzer identifies syntactic references in YAML
definitions but does not perform runtime analysis.
A task change that alters its output results may affect downstream
tasks in ways the static analyzer cannot detect.
Combining static analysis with runtime trace data could provide
a more complete impact picture.

\paragraph{Single LLM provider.}
The current implementation is coupled to the Google Gemini API.
While the prompt design is provider-agnostic, the summarizer does
not support provider failover or cost optimization through model
routing.

\subsection{Threats to Validity}

\paragraph{Construct validity.}
Our case study reports system characteristics and illustrative
examples rather than controlled experimental measurements.
The examples are representative of actual promotion patterns but
were selected to demonstrate the framework's capabilities.

\paragraph{External validity.}
The framework is designed for and evaluated within a single
platform's release catalog.
Generalization to other Tekton-based platforms should be
straightforward, as the task-pipeline structure is a Tekton
convention.
Generalization to non-Tekton CI/CD systems (e.g., Argo Workflows,
Jenkins) would require reimplementing the dependency analyzer
for the target system's pipeline definition format.

\paragraph{Internal validity.}
The LLM component introduces non-determinism.
We mitigate this through prompt engineering and temperature control
but cannot guarantee identical output across runs.

\section{Conclusion}
\label{sec:conclusion}

We presented a framework for AI-augmented release intelligence
that addresses the gap between user-facing release note generation
and internal engineering communication needs in cloud-native
delivery platforms.
By combining LLM-powered change summarization with static
task-pipeline dependency analysis and embedding both within the
CI/CD promotion workflow, the framework provides release engineers
with timely, structured, and actionable promotion reports.

The work opens several directions for future research.
Developing a benchmark for internal promotion report quality
would enable quantitative evaluation of summarization approaches
in this domain.
Extending the dependency analyzer with runtime trace integration
could capture transitive impacts invisible to static analysis.
Exploring multi-model ensembles or verification chains could
improve summary consistency without sacrificing naturalness.
Finally, broadening the framework to support additional CI/CD
platforms beyond Tekton would increase its applicability across
the cloud-native ecosystem.

\bibliographystyle{plain}
\bibliography{references}

\begin{thebibliography}{10}

\bibitem{smartnote2025}
Farbod Daneshyan, Runzhi He, Jianyu Wu, and Minghui Zhou.
\newblock {SmartNote}: An {LLM}-powered, personalised release note generator that just works.
\newblock {\em arXiv preprint arXiv:2505.17977}, 2025.
\newblock Accepted at FSE 2025.

\bibitem{forsgren2018accelerate}
Nicole Forsgren, Jez Humble, and Gene Kim.
\newblock {\em Accelerate: The Science of Lean Software and {DevOps}: Building and Scaling High Performing Technology Organizations}.
\newblock IT Revolution, 2018.

\bibitem{verlog2025}
Jiawei Guo, Haoran Yang, and Haipeng Cai.
\newblock {VerLog}: Enhancing release note generation for {Android} apps using large language models.
\newblock {\em Proceedings of the ACM SIGSOFT International Symposium on Software Testing and Analysis (ISSTA)}, 2025.

\bibitem{humble2010continuous}
Jez Humble and David Farley.
\newblock {\em Continuous Delivery: Reliable Software Releases through Build, Test, and Deployment Automation}.
\newblock Addison-Wesley Professional, 2010.

\bibitem{deeprelease2022}
Zhiqiang Jiang, Hui Liu, Zhenyu Niu, Lu~Zhang, and Gang Huang.
\newblock {DeepRelease}: Language-independent release notes generation from {Git} logs.
\newblock {\em Empirical Software Engineering}, 27(6), 2022.

\bibitem{konflux2024}
{Konflux CI Community}.
\newblock Konflux: An opinionated {Kubernetes}-native security-first software factory.
\newblock \url{https://konflux-ci.dev}, 2024.
\newblock Accessed: 2026-03-15.

\bibitem{releaseeval2025}
Qianru Meng, Zhaochun Ren, and Joost Visser.
\newblock {ReleaseEval}: A benchmark for evaluating language models in automated release note generation.
\newblock {\em arXiv preprint arXiv:2511.02713}, 2025.

\bibitem{moreno2014automatic}
Laura Moreno, Gabriele Bavota, Massimiliano Di~Penta, Rocco Oliveto, Andrian Marcus, and Gerardo Canfora.
\newblock Automatic generation of release notes.
\newblock {\em Proceedings of the 22nd ACM SIGSOFT International Symposium on Foundations of Software Engineering}, pages 484--495, 2014.

\bibitem{metagente2025}
Duc S.~H. Nguyen, Bach~G. Truong, Phuong~T. Nguyen, Juri Di~Rocco, and Davide Di~Ruscio.
\newblock Automated summarization of software documents: An {LLM}-based multi-agent approach.
\newblock {\em Automated Software Engineering}, 2025.

\bibitem{gemini2024}
Gemini Team, Rohan Anil, Sebastian Borgeaud, et~al.
\newblock Gemini: A family of highly capable multimodal models, 2024.
\newblock Google DeepMind Technical Report.

\bibitem{tekton2019}
{The Tekton Authors}.
\newblock Tekton: Cloud-native {CI/CD} pipelines.
\newblock \url{https://tekton.dev}, 2019.
\newblock Accessed: 2026-03-15.

\bibitem{logsage2025}
Weiyuan Xu, Juntao Luo, Tao Huang, Kaixin Sui, Jie Geng, Qijun Ma, Isami Akasaka, Xiaoxue Shi, Jing Tang, and Peng Cai.
\newblock {LogSage}: An {LLM}-based framework for {CI/CD} failure detection and remediation with industrial validation.
\newblock {\em arXiv preprint arXiv:2506.03691}, 2025.

\bibitem{agentdevel2026}
Di~Zhang.
\newblock {AgentDevel}: Reframing self-evolving {LLM} agents as release engineering.
\newblock {\em arXiv preprint arXiv:2601.04620}, 2026.

\end{thebibliography}

\end{document}